\documentclass[useAMS,usenatbib]{mn2e}
\usepackage{amssymb}
\usepackage{amsmath}
\usepackage{graphicx}
\usepackage{epstopdf}
\usepackage{array}
\usepackage{natbib}
\usepackage{caption}
\usepackage{threeparttable}
\usepackage{color}

\title[]{LAMOST J064137.77+045743.8: A New Binary of an A7-type Pulsating Subgiant and an M-type Red Dwarf}
\author[Y. H. Chen, C. M. Duan, B. K. Sun]{Y. H. Chen$^{1,2,3}$\thanks{E-mail: yanhuichen1987@126.com}, C. M. Duan$^{2}$, B. K. Sun$^{1}$\\
$^{1}$Institute of Astrophysics, Chuxiong Normal University, Chuxiong 675000, China\\
$^{2}$Faculty of Science, Kunming University of Science and Technology, Kunming 650093, China\\
$^{3}$International Centre of Supernovae (ICESUN), Yunnan Key Laboratory, Kunming 650216, China}

\begin{document}

\date{Accepted: }

\pagerange{\pageref{firstpage}--\pageref{lastpage}} \pubyear{}

\maketitle

\label{firstpage}

\begin{abstract}
With the progressive release of data from numerous sky surveys, humanity has entered the era of astronomical big data. Multi-wavelength, multi-method research is playing an increasingly crucial role. Binaries account for a substantial fraction of all stellar systems and research into binaries is of fundamental importance. LAMOST J064137.77+045743.8 has not yet been recorded in the SIMBAD astronomical database. We conducted a comprehensive analysis of LAMOST J064137.77+045743.8 using multi-band spectroscopic, astrometric, and photometric data. The low-resolution spectra from Large Sky Area Multi-Object Fiber Spectroscopic Telescope (LAMOST) suggest that LAMOST J064137.77+045743.8 is a binary consisting of an A7-type subgiant star ($T_{\rm eff}$ $\sim$ 7500\,K and log\,$g$ $\sim$ 3.9) and a cool red dwarf star. Astrometric data from Globe Astrometric Interferometers for Astrophysics support the binary speculation with a Renormalized Unit Weight Error metric value of 1.9. Additional flux observations in the infrared bands further corroborate the presence of a red dwarf companion. The i-band flare detected by the Zwicky Transient Facility (ZTF) photometric observations bolsters the interpretation of an M-type red dwarf companion. The radial velocity variations in the H$\alpha$ lines from LAMOST medium-resolution spectra and the light curves from ZTF both support the classification of the A7 subgiant as a pulsating star. The binary either has a long orbital period, a non-eclipsing binary orbit, or extremely shallow eclipses. Future asteroseismology studies will further probe the internal physics of the A7 subgiants. Research on binaries is incredibly fascinating.
\end{abstract}


\section{Introduction}

The advent of Einstein's mass-energy equation pointed the way for humanity to solve the problem of stellar energy sources. For over a century, despite rapid advancements in various fields of astronomy, the theory of stellar structure and evolution remains one of the most well-established areas in the discipline. The statistical distribution of stars in the Hertzsprung-Russell (H-R) diagram tells us that the vast majority of stars are main-sequence (MS) stars. Stars with a mass less than $\sim$2.2\,$M_{\odot}$ are classified as low-mass stars. In these stars, the central helium (He) core becomes electron-degenerate before it is ignited. Stars with a mass between $\sim$2.2\,$M_{\odot}$ and $\sim$9.0\,$M_{\odot}$ are classified as intermediate-mass stars. For these stars, the central carbon/oxygen (C/O) core is electron-degenerate prior to its ignition. Stars with a mass greater than $\sim$9.0\,$M_{\odot}$ are classified as massive stars. In massive stars, the central C/O core is non-degenerate before ignition. The mass thresholds that separate low-mass, intermediate-mass, and massive stars, based on the degeneracy of their cores at various evolutionary stages, are discussed in standard stellar astrophysics textbooks (see, e.g., Prialnik, 2009; Carroll \& Ostlie, 2017). The vast majority of low- and intermediate-mass stars end their lives as white dwarfs (WDs, Winget \& Kepler 2008). In contrast, massive stars terminate their evolution in a supernova explosion, which leaves behind a neutron star or a black hole. Stars constitute the bulk of celestial objects, and stellar physics is the cornerstone of astrophysics, making it a field of paramount importance.

Observational statistics of stars in the solar neighborhood (Duquennoy \& Mayor 1991; Raghavan et al. 2010), as well as simulations of stellar birth rates (Machida et al. 2005), both indicate that binaries account for a substantial fraction of all stellar systems. Research into binary systems is of fundamental importance. Recent years have witnessed astronomy's entry into a big data era, marked by an abundance of data continuously released from various sky survey telescopes. For the Early Data Release 3 (EDR3) of Globe Astrometric Interferometers for Astrophysics (GAIA, Gaia Collaboration et al. 2016), Gaia Collaboration et al. (2021) presented the Gaia Catalogue of Nearby Stars (GCNS) and reported 16,556 resolved binary candidates. Based on the radial velocity (Rv) data released by Sloan Digital Sky Survey (SDSS, York et al. 2000), Pourbaix et al. (2005) derived 675 possible new spectroscopic binary stars and orbits for 8 of them. Thanks to the large field of view and faint limiting magnitude of the Zwicky Transient Facility (ZTF, Bellm et al. 2019) DR2, Chen et al. (2020) classified 78,602 periodic variables, including $\sim$350,000 eclipsing binaries. Research on binaries is incredibly fascinating.

The Large Sky Area Multi-Object Fiber Spectroscopic Telescope (LAMOST) adopts an innovative active optics technique, giving it both a large aperture (3.6-4.9\,m) and a wide field of view (20 square degrees). The distributive parallel-controllable fiber positioning technique enables LAMOST accurately locate 4,000 astronomical objects simultaneously (Cui et al. 2012). LAMOST focuses on a number of contemporary cutting edge topics in astrophysics, including the first generation stars in the Galaxy, the formation and evolution history of galaxies, and the signatures of the distribution of dark matter and so on through the northern sky spectral survey (Zhao et al. 2012). From the pilot survey (Oct. 24, 2011 to Jun. 17, 2012) to the twelfth year survey (Sep. 18, 2023 to Jun. 3, 2024), LAMOST DR12 lwo resolution search (LRS) has released 12,605,485 spectra, including stars, galaxies, quasi-stellar objects, white dwarfs, cataclysmic variable stars, and so on. The LAMOST DR12 medium resolution search (MRS) has released 15,471,948 spectra, including 3,361,604 non time-domain data and 12,110,344 time-domain data. From Oct. 1, 2024 to Jun. 30, 2025, LAMOST DR13 has released 861,916 LRS spectra and 1,884,766 MRS spectra. The enormous number of celestial spectra provides us with abundant spectral resources for conducting astrophysical research. Qian et al. (2018) reported that by June 16, 2017, LAMOST had observed 3,196 EA-type eclipsing binaries, of which 2,020 had their atmospheric parameters precisely determined.

LAMOST J064137.77+045743.8 is a highly intriguing celestial object that came to my attention during my research on white dwarf spectra. Our study reveals that it is a binary system, which has not yet been cataloged in the SIMBAD astronomical database. In Sect. 2, we perform a predictive study for LAMOST J064137.77+045743.8 based on LRS and MRS spectra released by LAMOST. A supplementary verification for LAMOST J064137.77+045743.8 is conducted in Sect. 3 based on other spectroscopic surveys and imaging surveys. In Sect. 4, we derive a solid evidence of the companion star for LAMOST J064137.77+045743.8 based on data released by ZTF telescope. At last, we give a discussion and conclusions in Sect. 5.

\section{A predictive study for LAMOST J064137.77+045743.8 based on LRS and MRS spectra released by LAMOST}

\begin{figure}
\begin{center}
\includegraphics[width=8.8cm,angle=0]{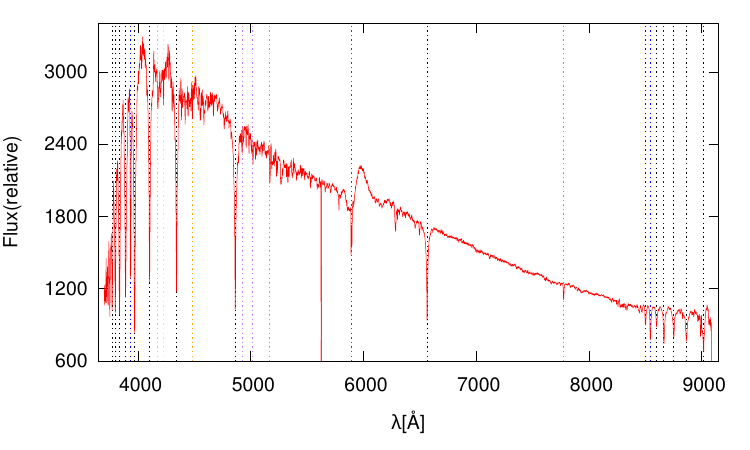}
\end{center}
\caption{LAMOST DR13 LRS spectrum for LAMOST J064137.77+045743.8 observed in Jan. 1, 2025.}
\end{figure}

\begin{table}
\begin{center}
\caption{The wavelength of absorption lines in angstrom identified in Fig. 1.}
\begin{tabular}{llllllllllll}
\hline
absorption lines                                                &wavelength[${\AA}$]              &color        \\
\hline
H9                                                              &\,3771                           &black        \\
H8                                                              &\,3798                           &black        \\
H$\eta$                                                         &\,3836.47                        &black        \\
H$\zeta$                                                        &\,3890.15                        &black        \\
H$\epsilon$                                                     &\,3971.19                        &black        \\
H$\delta$                                                       &\,4102.89                        &black        \\
H$\gamma$                                                       &\,4341.68                        &black        \\
H$\beta$                                                        &\,4862.68                        &black        \\
H$\alpha$                                                       &\,6564.61                        &black        \\
Paschen(n=14$\rightarrow$n=3)                                   &\,8598                           &black        \\
Paschen(n=13$\rightarrow$n=3)                                   &\,8665                           &black        \\
Paschen(n=12$\rightarrow$n=3)                                   &\,8750                           &black        \\
Paschen(n=11$\rightarrow$n=3)                                   &\,8863                           &black        \\
Paschen(n=10$\rightarrow$n=3)                                   &\,9015                           &black        \\
Fe II                                                           &\,4924                           &purple       \\
Fe II                                                           &\,5018                           &purple       \\
Fe II                                                           &\,5169                           &purple       \\
Mg II                                                           &\,4481                           &orange       \\
Ca II K                                                         &\,3934.78                        &blue         \\
Ca II triplet                                                   &\,8500.35,\,8544.44,\,8664.52    &blue         \\
Fe I                                                            &\,4176                           &grey         \\
Ca I                                                            &\,4227.92                        &yellow       \\
Na I                                                            &\,5896                           &brown        \\
O I                                                             &\,7775                           &green        \\
\hline
\end{tabular}
\end{center}
\end{table}

For the LAMOST LRS spectra, LAMOST J064137.77+045743.8 was observed six times in Feb. 18, 2012, Nov. 11, 2023, Nov. 16, 2023, Nov. 17, 2023, Dec. 29, 2024, and Jan. 1, 2025, with identified spectral types of A5, WD, A7, F0, A7, and A7 respectively. For the second spectrum, the signal to noise ratio (SNR) is 0.0 for the r, i, and z bands respectively. It could be easily mistaken for a WD, however, its spectrum contains an over abundance of metal lines, as shown in Fig. 1 and Table 1. The four most recent spectra are so similar that they can be brought into close agreement by a simple multiplicative scaling factor. Based on the four recently observed spectra from LAMOST, the derived effective temperature and gravitational acceleration are $T_{\rm eff}$ $\sim$ 7500\,K (7519.43 $\pm$ 25.08, 7527.08 $\pm$ 17.36, 7570.27 $\pm$ 18.41, and 7555.08 $\pm$ 18.15) and log\,$g$ $\sim$ 3.9 (3.931 $\pm$ 0.034, 3.865 $\pm$ 0.022, 3.931 $\pm$ 0.024, and 3.871 $\pm$ 0.023). The relatively weak and narrow absorption lines of Fe II and the Ca II K also indicates that the primary component of the spectrum is contributed by an A7 subgiant star. However, the Ca I line, Na I line, and Fe I line suggests the possible existence of a low-$T_{\rm eff}$ companion star. The Ca II triplet indicates that the cool companion star is likely a K or M type dwarf. Based on the LAMOST LRS spectra, LAMOST J064137.77+045743.8 is not a WD but most likely a combined system of an A7-type subgiant star and a K or M type red dwarf star. In addition, the broad emission around 6,000\,${\AA}$ should be instrument or reduction artefact.

\begin{figure}
\begin{center}
\includegraphics[width=8.8cm,angle=0]{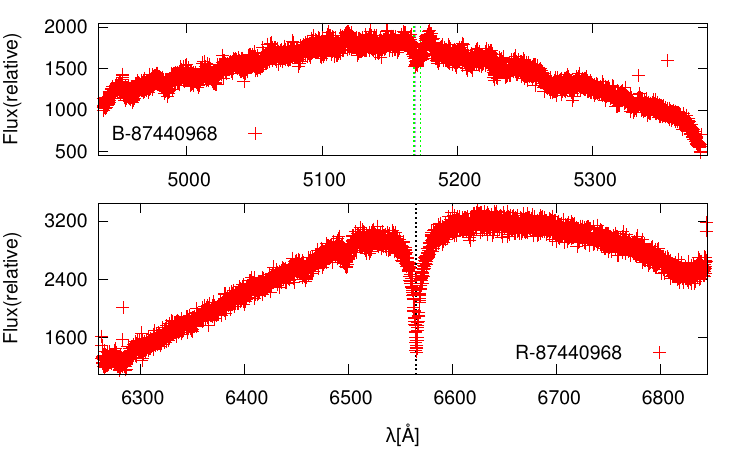}
\end{center}
\caption{LAMOST DR13 MRS spectra for LAMOST J064137.77+045743.8 observed in Feb. 16, 2025. The local modified Julian minute is 87,440,968.}
\end{figure}

For the LAMOST DR13 MRS spectra, the recent observational data, LAMOST J064137.77+045743.8 was observed over three days: Feb. 7, Feb. 16, and Mar. 6, 2025. For each day, there are three spectroscopic observations taken at 22-minute intervals, with each spectrum covering both the B-band (roughly from 4938\,${\AA}$ to 5380\,${\AA}$) and R-band (roughly from 6263\,${\AA}$ to 6843\,${\AA}$). The LAMOST DR13 MRS spectra also support the parameter values of $T_{\rm eff}$ $\sim$ 7500\,K (7508.13 $\pm$ 18.16 and 7695.40 $\pm$ 23.91) and log\,$g$ $\sim$ 3.9 (3.970 $\pm$ 0.020 and 3.845 $\pm$ 0.029) for the primary star. In Fig. 2, we show the B-band and R-band spectra with local modified Julian minute of 87,440,968. The H$\alpha$ absorption line (black) and the possible Fe II (purple) and Mg I lines (green) are marked as vertical dashed lines. We obtained the Rvs of 9 R-band spectra by fitting the H$\alpha$ absorption line with a Gaussian function, as shown in Table 2. The wavelength error at the center of the H$\alpha$ absorption line is on the order of 0.003\%. We have also used the Gaussian fitting method to study the value of Rv of a cataclysmic variable star IU Leo (Chen et al. 2025). The maximum Rv is 60.74\,km/s and the minimum Rv is 17.81\,km/s. With an interval of 22 minutes, the maximum difference of Rv can reach 33.8 km/s. Unlike cataclysmic variables, the variations in the spectroscopic Rv of LAMOST J064137.77+045743.8 should originate from the pulsations of the A7-type subgiant star. This is because binary systems with orbital periods of a few dozen minutes are dynamically unstable or exhibit intense mass transfer that generates emission lines. No emission lines were detected in either LAMOST LRS spectra or LAMOST MRS spectra.

\begin{table}
\begin{center}
\caption{The Rv values of 9 R-band spectra.}
\begin{tabular}{llllllllllll}
\hline
lmjm[minutes]                      &wavelength of  H$\alpha$[${\AA}$]    &Rv[km/s]                                       \\
\hline
87427981                           &6565.38                              &\,35.16                                        \\
87428003                           &6565.58                              &\,44.30                                        \\
87428025                           &6565.94                              &\,60.74                                        \\
87440924                           &6565.94                              &\,60.74                                        \\
87440946                           &6565.20                              &\,26.94                                        \\
87440968                           &6565.20                              &\,26.94                                        \\
87466798                           &6565.18                              &\,26.03                                        \\
87466820                           &6565.00                              &\,17.81                                        \\
87466842                           &6565.55                              &\,42.93                                        \\
\hline
\end{tabular}
\end{center}
\end{table}

\section{A supplementary verification for LAMOST J064137.77+045743.8 based on data released by GAIA, SkyMapper, 2MASS, and WISE}

\begin{table*}
\begin{center}
\caption{The apparent magnitude for LAMOST J064137.77+045743.8. The values of uvgriz, JHK, and w1w2w3w4 are released by SkyMapper, 2MASS, and WISE respectively. The center wavelength of uvgriz, JHK, and w1w2w3w4 are from Keller et al. 2007, Skrutskie et al. 2006, and Wright et al. 2010 respectively.}
\begin{tabular}{lcccccccccccccccccccccccccc}
\hline
filter             &u            &v           &g          &r           &i            &z           &J           &H          &K         &w1        &w2        &w3         &w4        \\
$\lambda$($nm$)    &\,355.0      &\,387.0     &\,497.0    &\,604.0     &\,771.0      &\,909.0     &\,1,250     &\,1,650    &\,2,160   &\,3,400   &\,4,600   &\,12,000   &\,22,000  \\
\hline
mag                &\,15.312     &\,14.441    &\,13.394   &\,13.085    &\,12.884     &\,12.736    &\,11.818    &\,11.544   &\,11.451  &\,11.339  &\,11.347  &\,12.286   &\,9.121   \\
err                &\,0.043      &\,0.022     &\,0.012    &\,0.013     &\,0.011      &\,0.016     &\,0.021     &\,0.020    &\,0.019   &\,0.012   &\,0.009   &\,0.412    &          \\
\hline
\end{tabular}
\end{center}
\end{table*}

\begin{figure}
\begin{center}
\includegraphics[width=8.8cm,angle=0]{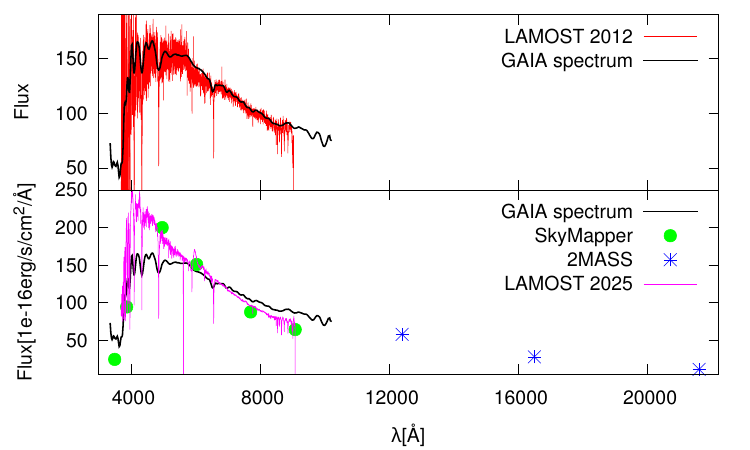}
\end{center}
\caption{Spectra data from LAMOST and Gaia together with flux data from SkyMapper and 2MASS.}
\end{figure}

Multi-wavelength studies are essential for a more comprehensive understanding of the target celestial object. LAMOST J064137.77+045743.8 was not observed by SDSS, but it was observed by Gaia, the SkyMapper telescope (Keller et al. 2007), the Two Micron All Sky Survey (2MASS, Skrutskie et al. 2006), and the Wide-field Infrared Survey Explorer (WISE, Wright et al. 2010). In Table 3, we show the apparent magnitude for LAMOST J064137.77+045743.8 in the uvgriz, JHK, and w1w2w3w4 bands respectively. The apparent magnitude in the near-infrared band indicates strong radiation in that band, which supports the speculation that the companion star is a K or M type red dwarf star.

In Fig. 3, we show the spectra data from LAMOST and Gaia XP (no Rvs Spectrum) together with flux data from SkyMapper, 2MASS, and WISE for LAMOST J064137.77+045743.8. The spectrum from LAMOST in 2025 is consistent with the flux data observed by the SkyMapper telescope. However, it is inconsistent with the XP spectrum from GAIA. These spectra were scaled by a factor during the comparison process. It is quite interesting that the spectrum from LAMOST in 2012 is consistent with the XP spectrum from GAIA. The spectra in the upper panel of Fig. 3 are likely significantly influenced by the companion star. In the GAIA archive website (https://gea.esac.esa.int/archive/), LAMOST J064137.77+045743.8 was fitted by a binary of $T_{\rm eff1}$ $\sim$ 6750\,K, log\,$g1$ $\sim$ 4.55 and $T_{\rm eff2}$ $\sim$ 5450\,K, log\,$g2$ $\sim$ 3.95. This was obtained from spectral fitting when the GAIA XP spectrum was significantly influenced by the companion star. We examined the Renormalized Unit Weight Error metric (RUWE, Lindegren et al. 2021) value and found it to be 1.9. This is significantly greater than the threshold of 1.2 (Krolikowski et al. 2021), which from an astrometric perspective indicates the presence of a binary star system.

\section{A solid evidence of the companion star for LAMOST J064137.77+045743.8 based on data released by ZTF telescope}

\begin{figure}
\begin{center}
\includegraphics[width=8.8cm,angle=0]{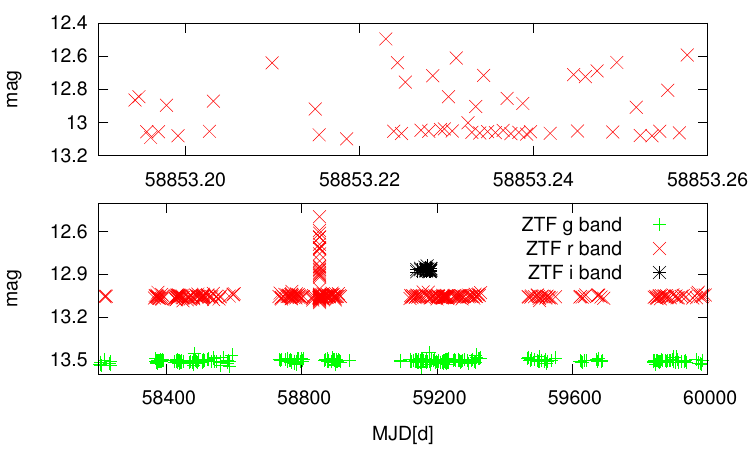}
\end{center}
\caption{Light curve of g, r, and i bands for LAMOST J064137.77+045743.8 from ZTF.}
\end{figure}

The Barbara A. Mikulski Archive for Space Telescopes (MAST, Hou et al. 2023) hosts data from numerous telescope missions and provides user-friendly access. We identified the target star by directly cross-matching LAMOST's DA-type WD data with MAST's archive data (CLT catalog). Time-domain data are essential for confirming binary systems. Unfortunately, LAMOST J064137.77+045743.8 was not observed by either the Kepler mission (Borucki et al. 2010) or the Transiting Exoplanet Survey Satellite (TESS, Ricker et al. 2014) in MAST archive. In MAST archive, the stellar parameters of LAMOST J064137.77+045743.8 are $T_{\rm eff}$ = 7657.52 $\pm$ 211.28\,K, log\,$g$ = 3.73 $\pm$ 0.11, radius = 3.01 $\pm$ 0.24\,$R_{\odot}$, mass = 1.79 $\pm$ 0.29\,$M_{\odot}$, luminosity = 28.10 $\pm$ 3.65\,$L_{\odot}$, and distance = 1848.24 $\pm$ 85.80\,pc. These parameters are self-consistent and support the classification of the primary star as an A7-type subgiant star.

Open-source astronomical big data is an invaluable asset for all humanity. Delighted to have found the light curve for LAMOST J064137.77+045743.8 on the ZTF website. The observations are from Mar. 27, 2018 to Feb. 18, 2023, covering 1,789 days, as shown in Fig. 4. The ZTF telescope employs a filter system comprising the g, r, and i bands (Dekany et al. 2020) for transient detection. There are 289, 366, and 27 data points for g, r, and i bands in the lower panel of Fig. 4 for LAMOST J064137.77+045743.8. In the r-band data, we detected an intermittent flare event. The brightness increased by 0.5 magnitudes, lasting for approximately 100 minutes, as shown in the upper panel of Fig. 4. The detected flare event provides strong evidence that the companion star is likely an M-type red dwarf, as such flares are caused by magnetic reconnection processes in the star's strong magnetic fields, which are generated within its convective zone.

Based on the light curve, five periods of 71.21, 196.4, 29.01, 6.016, and 27.33 days were derived from the ZTF website, with corresponding SNRs of 10.09, 9.503, 6.01, 5.749, and 5.716, respectively. These periods are also a manifestation of the pulsations in the A7-type subgiant star. No evidence of eclipses was found in over 1,789 days of photometric observations, as shown in Figure 4. This implies that either the binary orbital period is much longer than 1,789 days, or the system is highly inclined (showing very shallow or no eclipses). In the GAIA DR3 data, LAMOST J064137.77+045743.8 is located 1848\,pc away, making it extremely challenging to resolve into two point sources even for binary systems with very long orbital periods.

\section{A discussion and conclusions}

In recent years, with the continuous release and public availability of vast amounts of survey data, astronomical research has entered the era of big data. Multi-wavelength and multi-method studies have proven to be crucial in uncovering the true nature of research targets. In this paper, we serendipitously discovered LAMOST J064137.77+045743.8 while cross-matching data from the LAMOST data release with the MAST archive in search of pulsating white dwarfs. It has not yet been cataloged in the SIMBAD astronomical database. We have conducted studies on LAMOST J064137.77+045743.8 using multi-wavelength, multi-method approaches including spectroscopy, astrometry, and photometry.

Among the five valid low-resolution spectra released by LAMOST, four (3 A7-type and 1 F0-type) match the flux points measured by SkyMapper, while one (A5-type) aligns with the GAIA XP spectrum. Through detailed identification of the absorption lines in the four spectra, we found that in addition to the typical absorption lines of an A7-type star, there are also faint but distinct metal lines characteristic of low-temperature stars, such as Ca I, Na I, Fe I, and the Ca II triplet lines, as shown in Fig.1 and Table 1. Based on the effective temperature ($T_{\rm eff}$ $\sim$ 7500\,K) and surface gravity (log\,$g$ $\sim$ 3.9) values (Chen et al. 2022) derived from the spectroscopic data, along with detailed analysis of specific absorption lines, it is inferred that LAMOST J064137.77+045743.8 is likely a binary system composed of an A7-type subgiant and a K or M type red dwarf star. The flux measurements in the JHK bands from the 2MASS release and the W1, W2, W3, W4 bands from the WISE release likewise indicate the presence of a low-temperature star.

The RUWE parameter of GAIA for LAMOST J064137.77+045743.8 is 1.9, which indicates the presence of a binary star system (Krolikowski et al. 2021) from the astrometric perspective. The first spectrum for LAMOST J064137.77+045743.8 released by LAMOST is an A5-type star, consistent with the GAIA XP spectrum. We now know that this should be caused by flares from the red dwarf companion star. Multi-wavelength and multi-method studies yield more accurate and comprehensive binary star information than relying solely on GAIA data.

We are truly grateful for the open-source astronomical big data. The ZTF telescope has conducted 289, 366, and 27 valid observations in the g, r, and i bands, respectively, for LAMOST J064137.77+045743.8, spanning a time period of 1,789 days. In the r-band data, an intermittent stellar flare event with an apparent magnitude variation of 0.5 magnitudes and a duration of approximately 100 minutes was detected, as shown in the upper panel of Fig. 4. The stellar flare event provides strong evidence that the companion star is likely an M-type red dwarf. Therefore, LAMOST J064137.77+045743.8 is a binary with an A7-type subgiant primary star and an M-type red dwarf companion star. The absence of eclipses indicates that the system either has a long orbital period, a non-eclipsing binary orbit, or extremely shallow eclipses.

The radial velocity variations derived from the H$\alpha$ absorption lines in the LAMOST medium-resolution spectra originate from pulsations of the A7-type subgiant star for LAMOST J064137.77+045743.8. The ZTF website also provides five extracted pulsation periods for LAMOST J064137.77+045743.8. Jia et al. (2024) reported a largest catalog (2,254) of multimode $\delta$ Sct stars in the northern sky using the ZTF DR20. Asteroseismology of $\delta$ Sct stars is particularly fascinating, as it can even probe the size of the helium core in the stellar center (Chen et al. 217). In the future, we plan to conduct asteroseismology studies on this A7-type subgiant primary star to probe its internal physics.

\section{Acknowledgements}

Guoshoujing Telescope (the Large Sky Area Multi-Object Fiber Spectroscopic Telescope LAMOST) is a National Major Scientific Project built by the Chinese Academy of Sciences. Funding for the project has been provided by the National Development and Reform Commission. LAMOST is operated and managed by the National Astronomical Observatories, Chinese Academy of Sciences. This work is supported by the International Centre of Supernovae, Yunnan Key Laboratory (No. 202302AN36000101) and the Yunnan Provincial Department of Education Science Research Fund Project (No. 2024J0964).

\section*{Data availability}

The data underlying this article are available in the article and in its online supplementary material.

\label{lastpage}


\begin{thebibliography}{99}

\bibitem[\protect\citeauthoryear{Bellm}{2019}]{b1} Bellm E. C., Kulkarni S. R., Graham M. J., et al., 2019, PASP, 131, 018002
\bibitem[\protect\citeauthoryear{Borucki}{2010}]{b1} Borucki W. J., Koch D., Basri G., et al., 2010, Sci, 327, 977
\bibitem[\protect\citeauthoryear{Carroll}{2017}]{b1} Carroll B. W., Ostlie D. A., 2017, An Introduction to Modern Astrophysics, 2edn. Cambridge Univ. Press, Cambridge
\bibitem[\protect\citeauthoryear{Chen}{2020}]{b1} Chen X. D., Wang S., Deng L. C., et al., 2020, ApJS, 249, 18
\bibitem[\protect\citeauthoryear{Chen}{2017}]{b1} Chen X. H., Li Y., Lin G. F., et al., 2017, ApJ, 834, 146
\bibitem[\protect\citeauthoryear{Chen}{2025}]{b1} Chen Y. H., Duan C. M., \& Shu H., 2025, submitted to RAA
\bibitem[\protect\citeauthoryear{Chen}{2022}]{b1} Chen Y. H., Li G. W., \& Shu H., 2022, RAA, 22, 055008
\bibitem[\protect\citeauthoryear{Cui}{2012}]{b1} Cui X. Q., Zhao Y. H., Chu Y. Q., et al., 2012, RAA, 12, 1197
\bibitem[\protect\citeauthoryear{Dekany}{2020}]{b1} Dekany R., Smith R. M., Riddle R., et al., 2020, PASP, 132, 038001
\bibitem[\protect\citeauthoryear{Duquennoy}{1991}]{b1} Duquennoy A. \& Mayor M., 1991, A\&A, 248, 485D
\bibitem[\protect\citeauthoryear{GAIA}{2016}]{b1} GAIA Collaboration, Prusti T., de Bruijne J. H. J., Brown A. G. A., et al., 2016, A\&A, 595, A1
\bibitem[\protect\citeauthoryear{GAIA}{2021}]{b1} GAIA Collaboration, Smart R. L., Sarro L. M., Rybizki J., et al., 2021, A\&A, 649, A6
\bibitem[\protect\citeauthoryear{Hou}{2023}]{b1} Hou W., Luo A. L., Dong Y. Q., et al., 2023, ApJ, 165, 148
\bibitem[\protect\citeauthoryear{Jia}{2024}]{b1} Jia Q., Chen X. D., Wang S., et al., 2024, ApJS, 273, 7
\bibitem[\protect\citeauthoryear{Keller}{2007}]{b1} Keller S. C., Schmidt B. P., Bessell M. S., et al., 2007, PASA, 24, 1
\bibitem[\protect\citeauthoryear{Krolikowski}{2021}]{b1} Krolikowski D. M., Kraus A. L., \& Rizzuto A. C., 2021, AJ, 162, 110
\bibitem[\protect\citeauthoryear{Lindegren}{2021}]{b1} Lindegren L., Klioner A., Hern$\acute{a}$ndez J., et al., 2021, A\&A, 649, A2
\bibitem[\protect\citeauthoryear{Machida}{2005}]{b1} Machida M. N., Matsumoto T., Hanawa T., et al., 2005, MNRAS, 362, 382
\bibitem[\protect\citeauthoryear{Pourbaix}{2005}]{b1} Pourbaix D., Knapp G. R., Szkody P., et al., 2005, A\&A, 444, 643
\bibitem[\protect\citeauthoryear{Prialnik}{2009}]{b1} Prialnik D., 2009, An Introduction to the Theory of Stellar Structure and Evolution, Cambridge Univ. Press, Cambridge
\bibitem[\protect\citeauthoryear{Qian}{2018}]{b1} Qian S. B., Zhang J., He J. J., et al., 2018, ApJS, 235, 5
\bibitem[\protect\citeauthoryear{Raghavan}{2010}]{b1} Raghavan D., Mcalister H. A., Henry T. J., et al., 2010, ApJS, 190, 1
\bibitem[\protect\citeauthoryear{Ricker}{2014}]{b1} Ricker G. R., Winn J. N., Vanderspek R., et al., 2014, SPIE, 9143, 20
\bibitem[\protect\citeauthoryear{Skrutskie}{2006}]{b1} Skrutskie M. F., Cutri R. M., Stiening R., et al., 2006, ApJ, 131, 1163
\bibitem[\protect\citeauthoryear{Winget}{2008}]{b1} Winget D. E., Kepler S. O., 2008, ARA\&A, 46, 157
\bibitem[\protect\citeauthoryear{Wright}{2010}]{b1} Wright E. L., Eisenhardt P. R. M., Mainzer A. K., et al., 2010, ApJ, 140, 1868
\bibitem[\protect\citeauthoryear{York}{2000}]{b1} York D. G., Adelman J., Anderson J. E., et al., 2000, ApJ, 120, 1579
\bibitem[\protect\citeauthoryear{Zhao}{2012}]{b1} Zhao G., Zhao Y. H., Chu Y. Q., et al., 2012, RAA, 12, 723

\end{thebibliography}
\end{document}